**AC Susceptibility Studies in Fe doped $La_{0.65}Ca_{0.35}Mn_{1-x}Fe_xO_3$: Rare Earth Manganites.**

Wiqar Hussain Shah[1,2], S.K. Hasanain[3]

[1]Department of Physics, College of Science, King Faisal University,

Hofuf, 31982, Saudi Arabia

[2]Department of Physics, Federal Urdu University,
Islamabad, PAKISTAN

[3]Department of Physics, Quaid-i-Azam University, Islamabad, PAKISTAN

**Abstract:**

The effect of Fe substitution on Mn sites in the colossal magnetoresistive compounds $La_{0.65}Ca_{0.35}Mn_{1-x}Fe_xO_3$ with $0.01<x<0.1$ have been studied. A careful study in the magnetic properties has been carried out by the measurement of ac susceptibility. The temperature range of CMR is greatly broadened with the addition of Fe. Substitution of Fe induces a gradual transition from a metallic ferromagnetic with a high Curie temperature ($T_c$=270 K) to a ferromagnetic insulator with low $T_c$=79 K. Increased spin disorder and decrease of $T_c$ with increasing Fe content are evident. The variations in the critical temperature $T_c$ and magnetic moment show a rapid change at about 4-5% Fe. The effect of Fe is seen to be consistent with the disruption of the Mn-Mn exchange possibly due to the formation of magnetic clusters. An extra-ordinary behavior in the out of phase part ($\chi^{//}$) of ac susceptibility, characterized by double bump (shoulder), was observed around x=0.01 and 0.02. The shoulder in $\chi^{//}$ disappears at x≥0.04 Fe concentration. With increasing Fe concentration the $\chi^{//}$ peak shift to $T<T_{1/2}$ (mid point of the transition temperature) and becomes broader. The $\chi^{//}$ peak moves to 8 or 10K higher temperature on the application of a dc field, for 3 & 4% samples. We also see increasing low temperature dissipation in more strongly Fe doped samples i.e. increasing the Fe, leads to increased spin disorder and dissipation at low temperature. The effect of the dc field is discussed in terms of the suppression of spin fluctuations close to $T_c$. The same ionic radii of $Fe^{3+}$ and $Mn^{3+}$ cause no structure changes in either series, yet ferromagnetism has been consistently suppressed by Fe doping. Doping with Fe bypasses the usually dominant lattice effects, but depopulate the hopping electrons and thus weakens the double exchange. The results were explained in terms of the formation of magnetic clusters of Fe ions.

**INTRODUCTION:**

As is well known, $LaMnO_3$ has basically an anti-ferromagnetic arrangement of the spins at the transition metal sites and an insulating behavior. As the nearest neighboring compound of $LaMnO_3$, the perovskite iron oxide $LaFeO_3$ has similar transport and magnetic properties i.e., insulating and anti-ferromagnetic, however the hole doped $La_{1-x}A_xFeO_{3+N}$ continues to have an anti-ferromagnetic ground sate and remains insulating even at the maximum possible extent of hole doping. It is worth pointing out that $Mn^{3+}$ and $Fe^{3+}$ have almost identical ionic size [1]

To date, much of the exploration of the CMR materials has been done through doping of the La sites, which bring about the lattice effects, and ultimately influence the DE. Ghosh et al [2] studied the effect of doping in the compound $La_{0.70}Ca_{0.30}Mn_{1-x}Fe_xO_3$ with various transition elements on the CMR properties. They observed that the effects of local strain induced by the size mismatch between the dopants and the host lattice dominate the magnetic properties of such materials [2].

The study of the effects of Mn-site doping with other elements is expected to provide important clues concerning the mechanism of CMR as well as the exploring of novel CMR materials. During the past several years, there have been increasing reports on the effects of Mn-site elements substitution. It was found that doping on Mn sites with foreign elements generally decrease the Curie temperature $T_c$ and the insulator-metal (I-M) transition temperature $T_p$. However the exact effect depends on the nature of the doping element [3]. The doping element can be divalent (Mg, Cu, Ni and Zn), trivalent (Al, Fe, Ga, Co, Cr) and tetravalent (Ti and Sn). Among these doping elements, Fe has a special effect and attracts more attention [4]

The double exchange (DE), and consequently, the physical properties of these materials, is particularly susceptible to the lattice effects brought on by doping. Very strong lattice effects have been realized when the La ions are partially replaced by trivalent or divalent ions of different size. Any deviation from the ideal cubic perovskite structure can lead to either a reduction in Mn-O-Mn bond angle from $180^O$, or change in the bond strength, both directly affecting the DE [5].

An interesting way of modifying the properties of these systems is to dope the Mn site, which is the heart of the double exchange. Most investigation on Fe doped CMR materials have been performed on oxides whose un-doped compounds has a ferromagnetic transition below room temperature [6]. Simopoulos [7] studied Fe doping of the compounds $La_{1-x}Ca_xMnO_3$ by means of Mossbauer spectroscopy and observed that Fe is couples anti-ferromagnetically to its Mn neighbors. Ogale [8] studied the effect of doping in the compound $La_{0.70}Ca_{0.30}Mn_{1-x}Fe_xO_3$ and observed the occurrence of a localization-delocalization transition in the system at the critical concentration.

The doping mechanism for changing $Mn^{3+}/Mn^{4+}$ ratio in CMR materials is quite similar to that as followed for the High $T_c$ superconductors ($HT_cS$) to control effective Cu vacancy. Similar to $HT_cS$ compounds, both Cu and Mn can be substituted partially by other 3d metals like Co, Ni, Fe and Zn. In case of CMR compounds, Mn (3d) metal site substitution was also carried out and the results were explained in terms of $Mn^{3+}/Mn^{4+}$ ratio. $T_p$ decreases monotonically with increasing Fe concentration and disappeared for 18% Fe concentration. As expected, for most of the polycrystalline bulk materials, $T_p$ is not sharp for higher x, and paramagnetic to ferromagnetic transition takes place over a broad temperature range. The distortion of the Mn-O-Mn bond angle increases with x so that ferromagnetic double exchange (DE) interaction weakens [9]. At the transition temperature $T_p$, the CMR is highly sensitive to x, the doping concentration and δ, the oxygen deficiency in $R_{1-x}A_xMnO_{3\pm\delta}$ (R=La, Pr, Nd etc and A= Ca, Sr etc) samples. By changing x with fixed δ or vice versa, one essentially changes the $Mn^{3+}/Mn^{4+}$ ratio in these compounds. There exist a clear relation between $T_p$ and the amount of $Mn^{4+}$ ions.

An interesting way to modify the crucial $Mn^{3+}$-O-$Mn^{4+}$ network is to dope at the Mn site itself. It is well known that the magnetic behavior of this type of oxide depends on the $Mn^{4+}/Mn^{3+}$ pairs. Several authors studied the effect of Fe doping at the Mn site on the properties of perovskites. Righi et al [10] examined a 5% Fe-doped $La_{0.70}Ca_{0.30}MnO_3$ compounds and found a 50 K decrease in $T_c$ and 10-15% decrease in the average moment measured at 1 T. Pissas et al [11] studied the lightly (2%) Fe doped $La_{0.75}Ca_{0.25}MnO_3$ compound by Mossbauer spectroscopy and suggested the existence of ferromagnetic clusters in the sample with a size distribution. Ahn et al [5] observed the replacement of Mn by Fe favors insulating character and anti-ferromagnetism opposing the effects of DE. These authors argued that the nonparticipation of $Fe^{3+}$ in the DE results purely due to the electronic structure considerations. The absence of any signatures of $Fe^{4+}$ and $Fe^{5+}$ in the Mossbauer spectra in Fe doped samples further supports this conclusion. Consequently the Mn-O bond alone electrically active, with the carrier hopping occuring between the $Mn^{3+}$ and $Mn^{4+}$ ions. Cai et al [1] examined the properties of $La_{0.67}Ca_{0.33}Mn_{0.9}Fe_{0.1}O_3$ and confirm the occurrence of and competition between ferromagnetic and anti-ferromagnetic clusters in the system.

In order to obtained more information about the nature of the DE coupling, which is responsible for the ferromagnetic state, that is, the substitution of a small percentage of Mn atoms by another transition metal i.e. Fe. The samples have been studied with the aim of investigating the influence of the presence of a metal, which favors an anti-ferromagnetic coupling in the Mn-O layer. Another sample without Fe has also been studied to compare the results and to help in the understanding of the real effect of the Fe atoms.

We have previously reported the magnetic and transport behavior of Fe Doped $La_{0.65}Ca_{0.35}MnO_3$ compounds [12]. In accordance with the general observations we found it consistent decrease of the critical temperature $T_c$ with the Fe concentration. However we also observe that the critical temperature, magnetic moment and magnetoresistance effects showed a discontinuous change at around 4-5% Fe concentration. These changes were also reflected in the confinement length $1/\alpha$ obtained from the fit to the variable range hopping model. We have suggested along the lines of Ogale [8] that this critical concentration of Fe correspond to the case where the average Fe-Fe separation becomes comparable to the size of the small polarons. We further extended these studies with a view to observing the details of the magnetic transition by studying the in-phase $\chi'$ and out of phase $\chi''$ responses and their dc field dependence. Using a simple model we try to explain these changes particularly the change in terms of the variation of the electronic density of states with increasing Fe contents. It is suggested that drastic changes in the magnetic and transport properties may be reflective of the moment of the Fermi level from electron-like to hole-like region with increasing the Fe contents.

In this paper we discuss our investigation concerning the effect of Fe substitution on the prototype CMR materials $La_{0.65}Ca_{0.35}MnO_3$ by the in-Phase and out of Phase part of the ac-susceptibility with in a broad doping level 0.0″x″0.10. In this work, we have undertaken a study of the doping of the Mn site by Fe in $La_{0.70}Ca_{0.30}Mn_{1-x}Fe_xO_3$. Early studies have shown that in this Fe doping range, a direct replacement of $Mn^{3+}$ by $Fe^{3+}$ occurs [13]. Also unique to this doping, $Mn^{3+}$ and $Fe^{3+}$ have identical ionic sizes [14]. Consequently, the otherwise strong lattice effects can be bypassed, and the effects due to changes in the electronic structure become accessible.

In this work we present results, obtained from measurements of ac-susceptibility and their dc field effect on the influence of Fe doping (x″0.10) on the magnetic properties of the compound. There are two principle reasons why a study of this is important.

(i) Both Fe and Mn have similar ionic radii, so the crystalline structure is not modified by the addition of Fe.

(ii) The manganite $La_{0.65}Ca_{0.35}MnO_3$ possesses a large bandwidth, which is free of phenomenon such as electron phonon interactions characteristic of narrow bandwidth materials. Consequently, lattice effect may be ignored and effects due to the variation in electronic structure become accessible. Fe may therefore be used as a control parameter to vary the magnetic properties of these manganites.

Since Mn ion play key role in the electrical and magnetic properties of this system, which provides carriers, effective spins and local JT distortion, it is very worthwhile to investigate the influence of substitution on Mn sites with other elements. Here we present a study on the effects of the replacement of Mn by Fe in $La_{0.65}Ca_{0.35}MnO_3$. Because Mn and Fe have almost identical ionic radii, no strong lattice effects would be introduced by Fe substitution on Mn sites. Therefor the effects due to change in electrical structure become accessible. Our studies show that magnetic properties are much sensitive to the doping of Fe ion.

**Experiment:**

All samples reported in the present study were synthesized by standard solid state reaction procedure. The X-ray diffraction (XRD) measurements were carried out to confirm that single-phase materials had been prepared [12]. AC studies were conducted using a self-made ac probe with a split secondary (astatically wound) and a commercial lock-in amplifier [15].

AC studies were conducted in the range $h_{ac}$=5 Oe, f= 311 Hz and $0 < H_{dc} < 550$ Oe. For some selected samples we apply a dc field up to 1100 Oe. DC magnetic fields were obtained from a home made solenoidal magnet with the dc field direction parallel to that of ac field. Studies with superposed dc fields were conducted during warm up after the sample had been cooled down to 81K. Data were taken on bar shaped samples with masses of 11-18 mg. Most of the data shown is for the sample with the smaller mass.

**AC-susceptibility studies:**

All the compositions show the expected phase transition from paramagnetic to ferromagnetic phase. The width of the phase transition increases systematically with increasing Fe contents. All the $\chi^{/}$ vs T data exhibit a maximum after the sharp increase around $T_c$. The decline in the $\chi^{/}$ below the peak is very similar to the observed in the undoped ferromagnetic composition of ($La_{0.65}Ca_{0.35}MnO_3$) and has been discussed by us elsewhere.

The in-phase part of the susceptibility $\chi^{/}$ at $H_{dc}$=0 for Fe doped composition are studied in detail displayed in **Fig.1**. It is apparent from the Figure that the susceptibility $\chi^{/}$ decrease systematically with increasing Fe content down to x=0.04. However the discrepancies at lower temperature for x=≥0.04 are also apparent. All the composition goes from paramagnetic insulator phase to ferromagnetic metal phase at a particular temperature called the transition temperature $T_c$, and that particular temperature region are the transition region. As Fe is doped in to the sample both the ferromagnetic transition temperature $T_c$ and susceptibility are systematically lowered. The sharpness in the transition region from paramagnetic to ferromagnetic is also decreased. The decrease from the peak to the lowest temperature moment is also decreased. The increase in Fe concentration leads to increased spin disorder and dissipation at low temperature.

Typical data for the in phase part of the susceptibility $\chi'$, is shown in **Fig.1** for $h_{ac}$=5 Oe, f=311Hz and $H_{dc}$=0 Oe. One consistent feature to emerge from the $\chi'$ measurement at H=0 is the presence of the decrease in $\chi$<$T_c$ consistent with the dc magnetization behavior [12]. For H=0 there is roughly a 0-25% decrease in $\chi'$ below the maximum. The ac response $\chi'$ was studied in the more truly ferromagnetic $La_{0.65}Ca_{0.35}MnO_3$ and a very similar increase in $\chi'$ was observed with increasing temperature culminating in a very sharp drop at $T_c$. Thus both the dc [12] and ac magnetization show consistent behavior for both the compositions. As some maximum has to appear in the real part of ac susceptibility at the same temperature to that of the imaginary part for low Fe concentration. Obviously the paramagnetic to ferromagnetic transition temperature $T_c$ decreases with increasing x.

In the Fe doped sample no noticeable change of lattice parameters is observed, but it is also true that the ionic radius of $Fe^{3+}$ is smaller than that of $Mn^{3+}$/$Mn^{4+}$. Furthermore, it is logical to think the presence of $Fe^{3+}$ produces a distortion not detected in XRD diffraction pattern [12]. It is possible that this local distortion favors an anti-ferromagnetic arrangement catalyzed by the Fe ion.

A region of the paramagnetic phase contains ferromagnetic clusters due to Fe concentration, that increase in volume until the system undergoes a complete ferromagnetic transition. However, fluctuation behavior is observed in the clustering regime that a continuos over a large temperature region, and the phase stability behavior is more complex than previously reported [A. Tkachuk, K. Pogacki, D.E. Brown, B. Dabrowski, A.J. Fedro, C.W. Kimball, B. Payles, X. Xiong, Danill Rosenmann and B.D. Dunlap]. In summary we have observed suppression of ferromagnetism in ferromagnetic phase of $La_{0.65}Ca_{0.35}MnO_3$ by doping Fe on the Mn site. The usually dominant lattice effects have been bypassed due to identical size of $Mn^{3+}$ and $Fe^{3+}$.

### **DC field effect on $\chi'$**

With the application of the dc field H=550 Oe we observe several interesting features. In general the variation in $\chi'$ below the peak become more flat temperature independent, the magnitude of susceptibility decreases strongly, while the data begins to appears clearly separated in to two groups (a) x$\prec$0.04 and (b) x$\geq$(except for x=0.1 which is not completely the ferromagnetic transition). It is noticeable that the 4,5,7 and 8% Fe composition which here form the 2$^{nd}$ group not only have the strong decrease to the 1$^{st}$ group, but also show complete overlap for T<their respective $T_c$'s. It is also clear that these later compositions, the susceptibility variation does not show significant decline below $T_c$.

The data for $\chi'$ in the Zero field cooled (ZFC) state as shown in Fig.2 depicts the typical behavior seen in our range of $h_{ac}$ and $H_{dc}$. Firstly, there is a general decrease of $\chi'$ due to the applied dc field. For a field of 550 Oe there is a decrease of about 27-47% at the $\chi'$ maximum for different Fe composition. The decrease of the in phase part of $\chi'$ due to an applied dc field at low temperatures (T≤$T_c$) is not surprising. The other feature observed on application of the dc field is the flattening out of the low T decline of $\chi'$. For example for H=0 the decline down to 81K was seen earlier to be about 0-25%. However for H=550 Oe we find that this decrease in $\chi'$ is reduced to 0-8.9%. For an applied field of 660 Oe we find that the decline is completely eliminated, and the moments at T=81K and at the previous peak are equal. We have studied this effect in a number of compositions and find similar behavior. The extent of the field induced flattening varies with composition and also the qualitative features are changed to some extent. This field dependence is consistent with the dc behavior discussed earlier. The dc field while suppressing the overall response, due to change in the local susceptibility, acts to incorporate the smaller clusters into the extended cluster, suppressing the tendency for anti-ferromagnetic alignments.

The field dependence of susceptibility as a function of temperature is summarized in **Fig.3.** As observed in the undoped composition, the field dependence (the normalize change in susceptibility $\frac{\Delta\chi'}{\chi'}$ vs T) goes through a maximum at the corresponding $T_c$'s. This is also noticeable that the maximum decrease itself decreases with the increasing x initially. However for x≥0.04 there is again increase in the values of the field dependence e.g. for x=0.03 the maximum field effect is about 40% while for x=0.04 it is 52%.

We have conducted a study of the ac susceptibility with superimposed dc field 0<H<1100Oe in a series of *Fe* doped compounds, with x =0.01, 0.03, 0.04 and determined the changes in the susceptibility. The decline is demolished at 660 Oe filed from peak to lowest temperature. Our studies are particularly focussed on the effects at low fields, H<1100 Oe, as compared to previous studies. In **Fig.4** the magneto susceptibility (% change) of these compounds shows two interesting features. First it shows that the % age change decreases as x increases for 0,1,2,3 compositions. Then for x=4, 5 there is abrupt increase in % change means that more % age decrease in $\chi'$ on the application of the dc magnetic filed shown more sensitivity to the filed. Again beyond 5% Fe composition, the changes come down. The second interesting feature of these composition with the application of the magnetic field is that it shows a systematic decrease of $\chi'$ maximum with x till to 3% Fe composition. Then 4,5,7,8% Fe composition merge to same value at their respective $T_c$s. However 4 and 5% Fe composition are merge even when there is no dc magnetic field.

There is a local clustering of $Mn^{3+}$ ions around $Mn^{4+}$ ions due to strong ferromagnetic interactions. In the pure manganite (x=0) the local clustering of the $Mn^{3+}$ ions around $Mn^{4+}$ eventually leads to a ferromagnetic order of the long-range type. The existence of mixed exchange (i.e. presence of both ferromagnetic as well as AFM exchange), as well as diluting by the diamagnetic $Fe^{3+}$ ions, leads to disorder or canting of spins preventing complete ferromagnetic spin alignments. It is also possible that we have ferromagnetic clusters of finite dimensions which order below a certain temperature, albeit randomly, due to random anisotropy. The resulting magnetic order below $T_c$ therefore can be looked as diluted ferromagnetic created by the mixed exchange. To summarize we find that the Fe substitution break down the long rang ferromagnetic order seen in the pure manganites, replacing it with the mixed exchange.

## $1/\chi'$ analysis:

The presence of anti-ferromagnetic interactions was checked by measuring the Wiess constant, $\Sigma$, as a function of x, from a fit to reciprocal susceptibility in the paramagnetic phase. It is seen that $\Sigma$ decreases rapidly with increasing x. The typical data are shown in **Fig.B** for x=0.00, 0.01, 0.04, 0.05 and 0.10. We note that at low concentration in which the atoms are highly dilute, minimizing the probability of appreciable anti-ferromagnetic clusters. In the range $0.05 \supseteq x \supseteq 0.10$, at which some clusters are already present, the Fe clusters de-polarize the spins of their Mn neighbors, Hence we believe that moment, in this doping regime, is reduced by scattering off these de-polarized spins. This mechanism explains the large values of magneto-susceptibility of these samples at the temperature near the transition region. In this range, thermal fluctuations are very strong.

| Fe% | 0 | 1 | 2 | 3 | 4 | 5 | 7 | 8 | 10 |
|---|---|---|---|---|---|---|---|---|---|
| $T_c$ | 270 | 257 | 244 | 223 | 209 | 184 | 134 | 110 | 79 |
| $T_{\Sigma c}$ | 263 | 253 | 238 | 216 | 210 | 186 | 149 | 140 | - |

The values of the susceptibility show a gradual decrease on Fe substitution. $\Sigma_c$ also is reduced as x is increased, showing a general weakening of the ferromagnetic DE interaction and an increasing contribution of AFM interaction arising from superexchange. For the 4 and 5% Fe composition, $T_c \upsilon \Sigma$, however for other compositions there is significant difference between the two. The peak value of the susceptibility ($\chi_{ac}$) are lies in the mixed systems and it is the least for the compositions x=0.10. This lowering of the peak value of $\chi_{ac}$ most likely arises from the decrease in the magnetization $M_{dc}$ [12], which decrease on Fe substitution. The onset of the mixed type of ferromagnetic exchange interactions on Fe substitution therefore weakens the ferromagnetic order as well as lower the magnetic moment.

All the present data have shown that the DE interaction is fragile. Even a small amount of Fe ions seems to destroy the long-range ferromagnetic order, the metallic state. In a limited temperature range available in this measurement, the susceptibility above $T_c$ follows curie Wiess law. It can be seen that $\chi_{ac}$ decreases continuously as x increases and this arises because on Fe substitution we are decreasing the density of moment-carrying Mn ions.

**Out of phase response:**

As is well understood the out of phase response of a driven system (such as the spin system) is proportional to the losses taking place. These losses are typically large at and around a phase transition. The temperature variation of the out of phase response was studied for ac field amplitudes h=5 Oe, frequency f=311 Hz and dc field $0 \leq H \leq 550$ Oe. The out of phase part of the pure composition has a very sharp maximum peak and no shoulder. While x=0.01, 0.02 and 0.03% Fe composition have a shoulder before the main peak. As we increase the Fe concentration the peak become broad, the sharpness becomes less and also the peak value is decreased. The shoulder disappears at 4% and above this Fe concentration. Above the 5% Fe concentration there are no remarkable response of $\chi''$ at 5 Oe ac field and f=311Hz. The valley of the resistivity [12] matches with the peak of the $\chi''$ and also the shoulder in $\chi''$ matches with the resistivity peak for 0 and 1% Fe composition. Typical temperature dependence of the out of phase part, $\chi''$, for 1% Fe doped sample is shown in **Fig.5** for f=311Hz. and $h_{ac}$=5 Oe. There is a clear increase below about 260K, which in this case has a shoulder at, 257K; and maximum at 250K. The response then decreases continuously as T is lowered. A significant part of these data is the appearance of a shoulder. It will be seen that these features are retained at the application of dc field, albeit with somewhat less sharpness. We note that the response does not fall to zero below this broad peak and the shoulder but settles down to a finite value. This low temperature dissipation is in the same region where the in-phase part of susceptibility reaches a maximum. The continued presence of losses in the low temperature region is consistent with the picture of a gradual process of incorporation of small clusters into the main extended cluster, which generates the low temperature losses. However the first peak or shoulder in $\chi''$ appears (257K) for 1% Fe composition, somewhat below the maximum in resistivity and just at the mid of the susceptibility curve shown in **Fig.5**. For a system undergoing a magnetic transition the loss components are, according to the fluctuation dissipation theorem [2], proportional to the fluctuations in the magnetization. This loss component can be associated with the dynamical changes taking place in the critical region. As the dynamical correlations of the spin develop in this region, they are accompanied by energy losses and these are correspondingly reflected in the $\chi''$ behavior. This would correspond to the increase of $\chi''$ at 260K (1%Fe) and to the maximum around the mid-point of the transition, around 250K. The main peak in $\chi''$ however occurs close to $T_c$, the temperature where the magnetization M(T) shows no particular structure except for the slowing down of the changes in M. This suggests that the mechanisms responsible for the slowing down of the resistivity decrease are magnetic in origin and are associated with the increase in the loss component at 260K. It is notable that the loss component does not exhibit the global maximum at the middle of the magnetic transition, as is generally seen, but at a temperature (250K) significantly below it. It would appear from this that the losses continue to grow well up to the point where the magnetization continues to increase rapidly. Similar behavior is also observed in the sample with other Fe composition. We understand this to be originating in the continued rapid growth of the smaller clusters in this region, below $T_c$ itself. It is interesting to note that the µSR [3] and SANS data also shows the continued development of the small clusters down to a temperature about 20K below $T_c$. At lower temperatures the corresponding µSR amplitude decreases indicating the decrease in the number of smaller clusters.

### **DC field effect on $\chi''$**

We also discuss the dc field effect on the out of phase response. As is evident from the **Fig.6** the response is very strongly suppressed by the dc fields of the typical value of a few hundred Oe for 1% Fe composition. The effect is very strong at and close to $T_c$. The out of phase response is completely suppressed by fields.

It may be pointed out here that the suppression of $\chi''$ is much stronger than that of $\chi'$. The peak of $\chi''$ is shifted up to 8-10K higher temperature than the actual peak with the application of dc field upto 550 Oe in 3 and 4% Fe composition.

We understand the strong field dependence of the loss components at higher temperatures (T>$T_c$) as being due to the suppression of dynamical spin fluctuations in the transition region, as already mentioned in the context of the in-phase part of the susceptibility. With the application of the field the fluctuations are suppressed and the dc field immobilizes part of the spins leading to a decrease in the in-phase part as well. The suppression of the peak with field probably corresponds to the removal of local correlations due to the applied field.

**Conclusions:**

In summary the effect of Fe doping in $La_{0.65}Ca_{0.35}MnO_3$ on magnetic properties has been studied. An increase in Fe concentration in this material, decreased ferromagnetic ordering and produce large magneto-susceptibility near the peak. These results were explained as due to anti-ferromagnetic Fe clusters formation. Large value of magneto-susceptibility at low temperature and low fields (0″H″1100 Oe), attributed to spin polarized tunneling at the grain boundaries were observed.

The addition of the small amount of Fe to the $La_{0.65}Ca_{0.35}MnO_3$ perovskite system has a great influence in the magnetic behavior. Magnetic properties as curie temperature and the magnetic moment show a decrease in their values when Mn atom are substitutes by Fe atoms, probably due to anti-ferromagnetic coupling with the neighbors atoms. The doping of Fe results in a depletion in the number of hopping electrons and available hopping sites and hence weakens the double exchange interactions. This weakens causes the suppression of metallicity and the system is pushed towards the insulating side. As the concentration of Fe increases, the disorderness increases, causing the decreasing in the susceptibility. Since the long range ferromagnetism is weakened by the Fe, and the susceptibility is also affected and decreases as the Fe content increases in the system. Thus we can conclude that the DE and the resulting ferromagnetic state as seen in the CMR oxides is fragile and even a small amount of Fe alters the magnetism.

Our studies of the magnetic behavior are consistent with previous studies in the sense of showing a general decrease of $T_c$ with increasing *Fe* content. In view of the reports on these doped materials regarding the formation of spin clusters, it seems reasonable to argue that the effect of the field is to decrease the spin disorder, both above and below $T_c$, thereby increasing the spin correlations and the conductivity. At $T_c$ where the correlations are not fully established, the field apparently helps stabilize the ferromagnetic correlations in the doped materials. Our data are consistent with the development of an extended cluster or region where the critical fluctuations and their related losses are strongly suppressed on the application of dc fields. We find that the pattern of the losses is consistent with the rapid growth of clusters probably upto some temperature and continued presence of these smaller clusters well below this temperature. The effects of the dc field on the susceptibilities, $\chi^{/}$ and $\chi^{//}$, are explainable within the context of the suppression of fluctuations, away and close to the critical region. What is remarkable however is the very strong suppression of the out of phase part even for moderate dc fields ~550's of Oe. It is well known that DE mediates ferromagnetism and metallic conduction. The transport and magnetic result shown above clearly demonstrate that the partial replacement of Mn by Fe favors insulating and AF behavior, opposing the effect of double exchange. Since Fe doping is the direct replacement of $Mn^{3+}$ by $Fe^{3+}$, the experimental results suggests that the sites that are now occupied by $Fe^{3+}$ can no longer effectively participate in the DE process. The mechanism that $Fe^{3+}$ terminate the DE process arises purely from the electronic structure of the materials.

**Figure Captions**

Figure 1: Temperature dependence of the real part of the susceptibility for $La_{0.65}Ca_{0.35}Mn_{1-x}Fe_xO_3$ with $0.01<x<0.1$, taken ZFC (warming) curves at 5 Oe ac probing field at a fixed frequency of 131 Hz. The symbols used in different curves for different composition are ν , , , ,♦, +, ×,* and   for x= 0.00, 0.01, 0.02, 0.03, 0.04, 0.05, 0.07, 0.08 and 0.10 respectively.

Figure 2: Temperature dependence of the real part of the susceptibility for $La_{0.65}Ca_{0.35}Mn_{1-x}Fe_xO_3$ with $0.01<x<0.1$, taken ZFC (warming) curves at 5 Oe ac probing field with the superimposed dc field up to 550 Oe at a fixed frequency of 131 Hz. The symbols used in different curves for different composition are ν , , , ,♦, +, ×,* and   for x= 0.00, 0.01, 0.02, 0.03, 0.04, 0.05, 0.07, 0.08 and 0.10 respectively.

Figure 3: Variation of the percentage change of the in-phase part of the susceptibility vs temperature. The symbols used for different Fe compositions are: ν for x=0.01,   for x=0.02,   for x=0.04,   for x=0.05, ♦ for x=0.07 and + is for x=0.08.

Figure 4: Fe Concentration dependencies of the percentage change in the real part of the magneto-susceptibility at the peak value.

Figure B: Temperature dependence of Curie-Wiess law, $1/\chi'$, taken ZFC (warming curves) at 5 Oe ac probing field at a frequency of 131 Hz, for different Fe composition. The symbols used are ν , , , ,♦ for x= 0.00, 0.01, 0.04, 0.05 and 0.10 respectively.

Figure 5: Temperature dependence of the real part, imaginary part and resistivity in zero dc magnetic filed for 1% Fe composition. The data have been normalized to unity.

Figure 6: The dc field (H=550 Oe) effects are shown on the imaginary part of the susceptibility for 1% Fe composition.

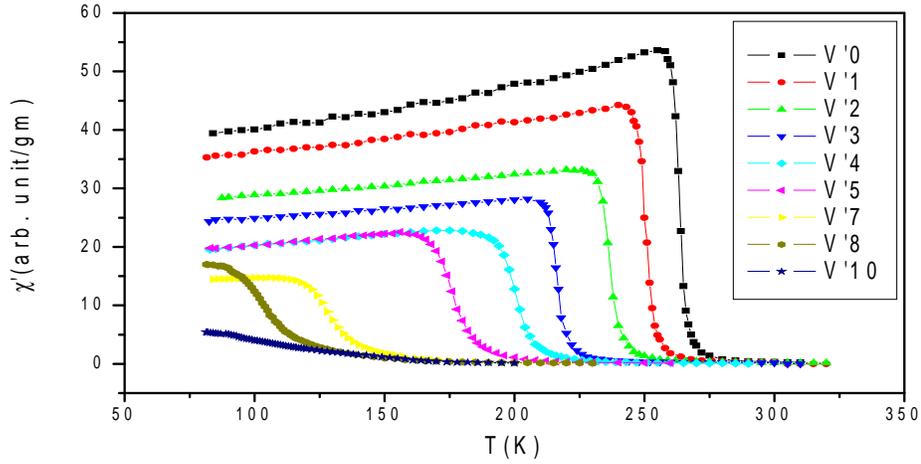

Figure 1: Temperature dependence of the real part of the susceptibility for $La_{0.65}Ca_{0.35}Mn_{1-x}Fe_xO_3$ with $0.01<x<0.1$, taken ZFC (warming) curves at 5 Oe ac probing field at a fixed frequency of 131 Hz. The symbols used in different curves for different composition are ν , , , ,♦, +, ×,* and for x= 0.00, 0.01, 0.02, 0.03, 0.04, 0.05, 0.07, 0.08 and 0.10 respectively.

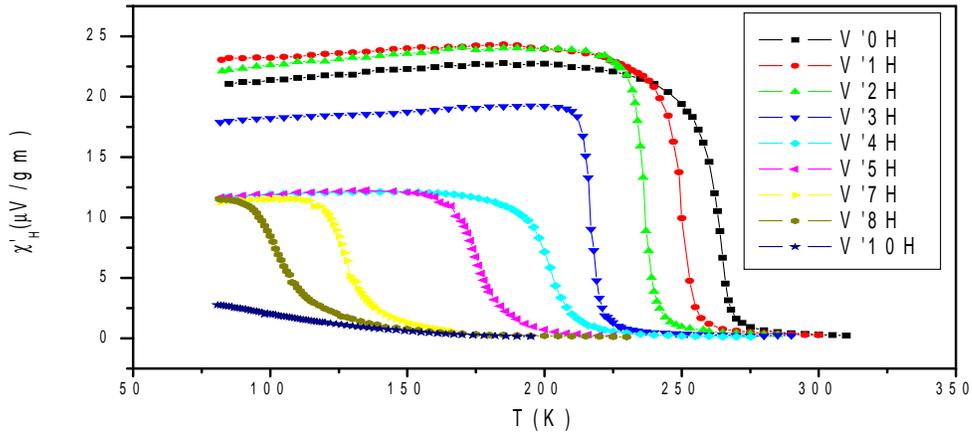

Figure 2: Temperature dependence of the real part of the susceptibility for $La_{0.65}Ca_{0.35}Mn_{1-x}Fe_xO_3$ with $0.01<x<0.1$, taken ZFC (warming) curves at 5 Oe ac probing field with the superimposed dc field up to 550 Oe at a fixed frequency of 131 Hz. The symbols used in different curves for different composition are ν , , , ,♦, +, ×,* and for x= 0.00, 0.01, 0.02, 0.03, 0.04, 0.05, 0.07, 0.08 and 0.10 respectively.

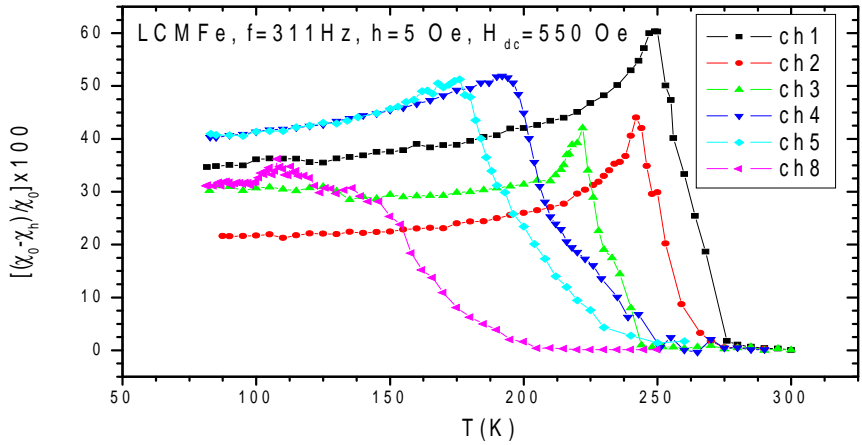

Figure 3: Variation of the percentage change of the in-phase part of the susceptibility vs temperature. The symbols used for different Fe compositions are: ν for x=0.01, for x=0.02, for x=0.04, for x=0.05, ◆ for x=0.07 and + is for x=0.08.

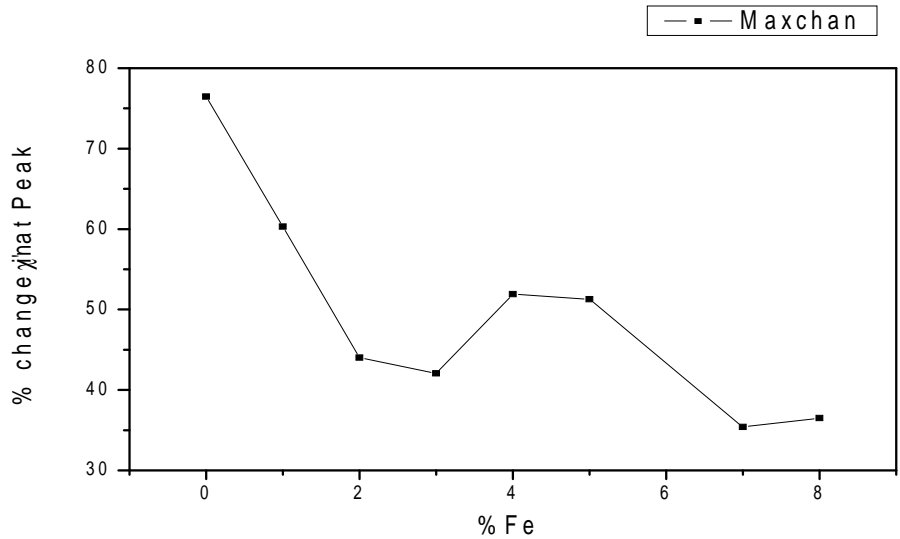

Figure 4: Fe Concentration dependencies of the percentage change in the real part of the magneto-susceptibility at the peak value.

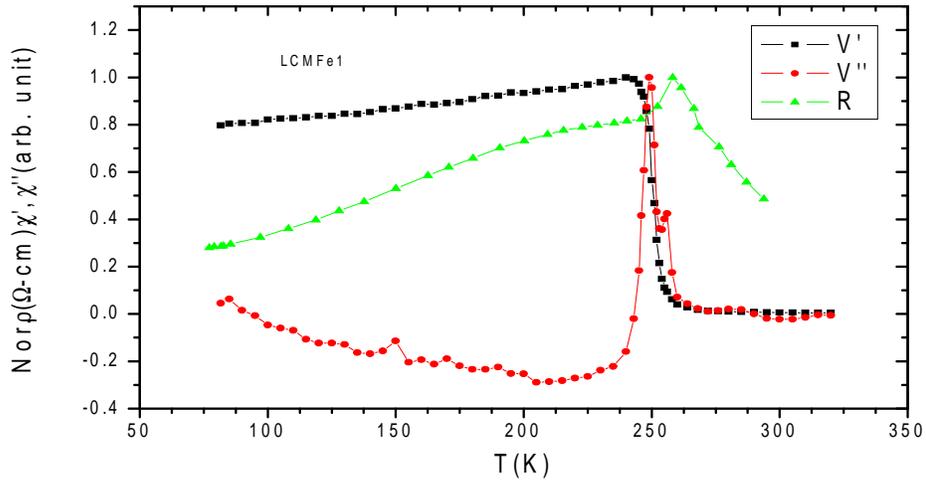

Figure 5: Temperature dependence of the real part, imaginary part and resistivity in zero dc magnetic field for 1% Fe composition. The data have been normalized to unity.

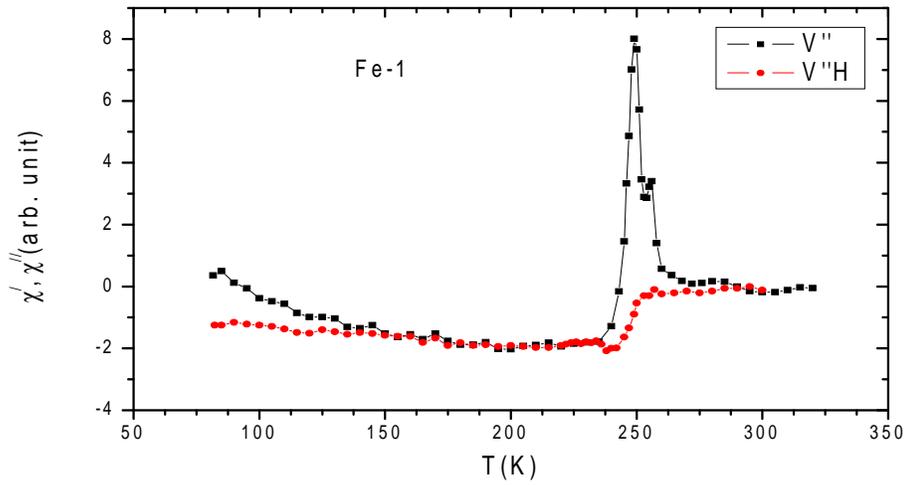

Figure 6: The dc field (H=550 Oe) effects are shown on the imaginary part of the susceptibility for 1% Fe composition.

16